\documentclass[12pt]{iopart}
\usepackage{graphicx}
\usepackage{latexsym}
\usepackage{amssymb}
\usepackage{bm}

\newcommand{\f}{\begin{equation}}
\newcommand{\ff}{\end{equation}}
  \def\msol{\ifmmode M_\odot\else$M_\odot$\fi}
  
\begin{document}

%\title{Applications of Numerical Relativity to the study of Early Universe Cosmology}
\title{Numerical Relativity as a tool for studying the Early Universe}

\author{David Garrison}
\address{Physics Department, University of Houston Clear Lake, Houston, Texas 77058}
\ead{garrison@uhcl.edu}

%\date{\today}

\begin{abstract}
Numerical simulations are becoming a more effective tool for conducting detailed investigations into the evolution of our universe.  In this article, we show how the framework of numerical relativity can be used for studying cosmological models.  The author is working to develop a large-scale simulation of the dynamical processes in the early universe.   These take into account interactions of dark matter, scalar perturbations, gravitational waves, magnetic fields and a turbulent plasma.  The code described in this report is a GRMHD code based on the Cactus framework and is structured to utilize one of several different differencing methods chosen at run-time.  It is being developed and tested on the University of Houston's  Maxwell cluster.
\end{abstract}

%\keywords{Cosmology: Numerical Relativity: Relativistic Plasma}


\maketitle

\section{Introduction}

Our knowledge of how the universe evolved comes primarily from observations of large structures such as stars, galaxies, clusters and super-clusters of galaxies as well as from observations of the Cosmic Microwave Background (CMB) radiation.  Based on these observations, the standard model of cosmology was developed during the mid to late twentieth century.  Some elements of this model include the existence of primordial metric perturbations, magnetic fields and an early universe filled with a nearly homogenous and isotropic plasma \cite{mukhanov}.  The perturbed Friedmann-Robertson-Walker (FRW) metric, which describes the space-time curvature of the early universe, takes the following form, 

\f
ds^{2} = -(1 + 2 {\phi}) dt^{2} + \omega_{i} {dt} {dx^{i}} + a(t)^{2} {\ } [  ( ( 1 + 2 {\ } \psi) {\ } \delta_{ij} + h_{ij} ) dx^{i} dx^{j}].
\label{eq:pert}
\ff

Here a(t) is the scale factor and $\phi$, $\psi$, $\omega_{i}$, and $h_{ij}$ are the scalar, vector and tensor perturbation terms.  Many cosmological models relate density fluctuations and variations in the CMB to perturbations in the FRW metric at the time of recombination.  These perturbations start off small and grow as a power-law with time as the competing forces of universal expansion and gravitational attraction affect their growth \cite{mukhanov}.  Work by Kodama and Sasaki \cite{kodama}, Sachs and Wolfe \cite{sachs} and Mukhanov, Feldman and Brandenberger \cite{mukhanov} all showed analytically how metric perturbations could cause density perturbations in a hydrodynamic fluid.

Recently, beyond the standard model cosmological theories~\cite{Kahniashvili:2008pf} have suggested that primordial fluids and fields are potential sources of observable gravitational waves. This realization opens up exciting new possibilities, making primordial gravitational radiation an important source of information about the early universe.  Taken together, this leads to the idea that there was a dynamical interaction between matter, electromagnetic and gravitational fields in the early universe that affected the evolution of our universe.  Signatures of these interactions may still be observable today.  The objective of this paper is to show how the tools of numerical relativity can be used to study such an interaction and to introduce a computer code written for this purpose.

This project uses the framework of numerical relativity to develop a computational laboratory to study the evolution of the early universe.  The initial focus of this work is on deriving the spectrum of gravitational waves produced by relativistic turbulence in the early universe.  Future work may involve a more advanced study of how this gravitational wave spectrum is effected by the presence of dark matter or a pre-existing primordial gravitational wave field.  Numerical relativity has been used for years to study the collisions of compact objects such as black holes and neutron stars and to predict the spectrum of the gravitational waves produced by their interactions.  We propose to use this tool to provide detailed studies of cosmological events that may also one-day be observable using either gravitational or conventional astronomy.  In cosmology, numerical simulations are capable of providing more detail than the analytic calculations that have been performed to date.  For example, a modern General Relativistic Magneto-HydroDynamic (GRMHD) code is capable of performing such simulations by evolving both the spacetime and plasma field dynamically and can therefore represent chaotic processes such as turbulence more accurately than analytic calculations.  This paper is a summary of the techniques used to develop such a simulation.

\section{Development of Initial Conditions}

Every numerical simulation consists of three parts: initial data, numerical evolution and data analysis.  In this section, we will focus on the initial data needed for cosmological studies. 

\subsection{Background and Hypotheses}

Several different mechanisms for producing primordial gravitational waves have been identified and studied by researchers.   These include quantum fluctuations during inflation, bubble wall motion and collisions during phase transitions, cosmological magnetic fields, oscillating classical fields during reheating, cosmological defects and plasma turbulence~\cite{Caprini:2006jb,Caprini:2009yp,Dolgov:2002ra,Gogoberidze:2007an,Kahniashvili:2005mp,Kahniashvili:2005qi,Kahniashvili:2006hy,Kahniashvili:2008pf,Kahniashvili:2008pe,Kahniashvili:2009mf,Nicolis:2003tg}.  We assume that the early universe was in a metastable state during a first order phase transition.  The false vacuum was separated from the true vacuum by a potential barrier or a scalar field.  Quantum tunneling occurred across the barrier in finite regions of space resulting in true vacuum bubbles inside the false vacuum phase.  As the universe expanded and cooled, the energy difference between the false vacuum and true vacuum got larger, making the phase transition more probable.  Eventually, the probability of nucleating one critical bubble per Hubble time became high enough to cause the phase transition to begin.  This defined the transition temperature, which is believed to be about 1 TeV~\cite{Kahniashvili:2008pf}.  The nucleated bubbles expanded and collided, eventually filling the whole universe.  The collision of two or more of these bubbles broke spherical symmetry and released some of their energy as gravitational waves.  Since the expansion of the bubbles was accompanied by macroscopic motions in the cosmic matter field, the collision of these bubbles also resulted in the anisotropic  stirring of the field.  This caused turbulent motions which provided a primary source of gravitational waves for this research.  

\subsubsection{Primordial Magnetic Fields}

Magnetic fields are believed to have played a large part in the dynamics of the universe's evolution.  Little is known about the existence of magnetic fields in the early universe. There are no direct observations of primordial magnetic fields.  Theories also disagree on the amplitude of primordial magnetic fields. There are currently several dozen theories about the origin of cosmic magnetic fields~\cite{battaner,grasso}.  The main reason that we believe that primordial magnetic fields existed is because they may have been needed to seed the large magnetic fields observed today.   Most theories of cosmic magnetic field generation fall into one of three categories~\cite{battaner,Dolgov:2001,grasso}: 1) magnetic fields generated by phase transitions; 2) electromagnetic perturbations expanded by inflation; and 3) turbulent magnetofluid resulting in charge and current asymmetries.

Most models calculate the magnitude of primordial magnetic fields by starting with the observed strength of galactic or intergalactic magnetic fields and calculating how this field should have been amplified or diffused by external effects such as the galactic dynamo and expansion of the universe~\cite{battaner,grasso}.  A major problem is that there doesn$\textquoteright$t appear to be a universal agreement of how efficiently a galactic dynamo could have strengthened seed magnetic fields.  Estimates of the strength of these seed fields can vary by tens of orders of magnitude.  Seed magnetic fields produced during Inflation are predicted to have a current strength somewhere between $10^{-11}$ G and $10^{-9}$ G on a scale of a few Mpc~\cite{battaner,grasso,hoyle}.  Magnetic seed fields generated by phase transitions are believed to be less than $10^{-23}$ G at galactic scales~\cite{battaner,grasso}.  Some turbulence theories imply that magnetic fields were not generated until after the first stars were formed therefore requiring no magnetic seed fields~\cite{battaner}.  

Given how little is understood about primordial magnetic fields and the general lack of agreement among theoretical predictions, it seems clear that the existence of primordial magnetic fields can neither be confirmed or ruled out.  It seems that the best we can do is set an upper limit on the strength of primordial magnetic fields and utilize this limit as a starting point in developing models of cosmic turbulence.  Observations of the CMB limit the intensity of the magnetic seed fields to a current upper limit  of $10^{-9}$ G~\cite{battaner,grasso,hoyle,murayama}.

It is well known that gravitational waves can interact with a magnetofluid in the presence of a magnetic field.  Work by Duez et al~\cite{duez2} showed how gravitational waves can induce oscillatory modes in a plasma field if magnetic fields are present.  Work by Kahniashvili and others ~\cite{Kahniashvili:2005mp,Kahniashvili:2005qi,Kahniashvili:2006hy,Kahniashvili:2008pf,Kahniashvili:2008pe,Kahniashvili:2009mf} have shown how a turbulent plasma can yield gravitational waves.  The result may be a highly nonlinear interaction as energy is transferred from the fluid to the gravitational waves and back.

\subsubsection{Turbulence in the Early Universe}

Turbulence provides a particularly interesting GW source because it is not well understood analytically.  This turbulence is a natural result of dynamics of the early universe resulting from bubble wall collisions and other chaotic events during the  first order phase transitions. Analytic work done to date \cite{Caprini:2006jb,Kahniashvili:2008pf,Kahniashvili:2008pe,Kahniashvili:2009mf,Nicolis:2003tg} summarizes the dynamics of the phase transitions using two quantities, $\alpha$ and $\beta$. $\alpha$ is traditionally defined as the ratio of false vacuum energy and plasma thermal energy density.  This provides a measure of the transition strength.  If $\alpha$ is much less than one, the transition is very weak.  If $\alpha$ is larger than unity, the transition is very strongly first order.  $\beta$ is the rate of variation of the nucleation rate at the transition time.  It fixes the time scale of the phase transition once the transition has begun.  After a time interval $\beta^{-1}$, the whole universe is converted to a true vacuum phase.  Therefore the turbulent stirring should only last $\beta^{-1}$.

The amount of gravitational waves emitted by bubble collisions and turbulence generated in the plasma are also determined from two quantities, $\kappa$ and $v_b$.  $\kappa$ is the fraction of vacuum energy transferred into fluid kinetic energy  and $v_b$ is the velocity of bubble wall expansion.  Bubble walls can propagate via two modes, detonation and deflagration \cite{Caprini:2006jb,Kahniashvili:2008pf,Kahniashvili:2008pe,Kahniashvili:2009mf,Nicolis:2003tg}.  For detonation, the bubble walls are thin compared to the radius and they propagate faster than the speed of sound.  This results in:  
\f
v_b(\alpha) = \frac{1 / \sqrt{3} + (\alpha^2 + 2 \alpha / 3)^{1/2} }{1 + \alpha }  \label{eq:vb}
\ff
\f
\kappa(\alpha) = \frac{1}{1 + 0.715 \alpha} \left[ 0.715 \alpha + \frac{4}{27} \sqrt{\frac{3 \alpha}{2}}  \right].
\ff
If the bubbles propagate by deflagration, the walls are thick  and have a lower energy density.  It is currently believed that for a relativistic plasma, the deflagration expansion mode is unstable so only the detonation modes will result.  

The number density of turbulent eddies within a Hubble radius should depend on $v_b$ and $\beta$.  The characteristic velocity perturbation of the turbulent fluid for the largest eddies at the stirring scale is given by:
\f
v_0 = \sqrt{ \frac{3 \kappa \alpha} {4 + 3 \kappa \alpha}}.  \label{eq:v_0}
\ff
For $\kappa \alpha$ $\approx$ 1, which corresponds to a strongly first order phase transition, $v_0$ is about 0.65 at the time of the electroweak phase transition.  We later use $v_0$ as the maximum velocity of fluid elements in our studies.  This velocity is randomized in amplitude and direction in order to simulate the initial conditions for turbulence.

\subsection{Computational Model}

In order to study the interaction of the plasma field and the background spacetime dynamically (or separately) our team has written and is testing/improving a GRMHD code implemented using the Cactus framework~\cite{garrison2}. We now describe the basic variables and equations that constitute this model.

\subsubsection{The Spacetime Evolution Model}

The spacetime metric can be written as: 
\f
ds^{2} = - N^{2} dt^{2} + \gamma_{ij} (\vec{x}, t)(dx^{i} + N^{i}dt)(dx^{j} + N^{j}dt).
\label{eq:adm}
\ff
Here $N$ is the lapse, $N^{i}$ is the shift vector and $\gamma_{ij}$ is the spatial 3-metric~\cite{bs065,duez1}.   For this work, 3-metric and its ``time-derivative'',  the extrinsic curvature, ``$K_{ij}$'' will be evolved using a strongly hyperbolic version of the BSSN formulation of numerical relativity~\cite{sarbach}.

\subsubsection{The General Relativistic Magnetohydrodynamic Model}
 
The fluid and electromagnetic fields of the GRMHD equations are developed from several well-known equations~\cite{harm}. They include the conservation of particle number, the continuity equation, the conservation of energy-momentum, the magnetic constraint equation and the magnetic induction equation. For a system consisting of a perfect fluid and an electromagnetic field, the ideal MHD stress-energy tensor is given by 
\begin{eqnarray}
T^{\mu\nu}  & = (\rho_0 h + b^{2})u^{\mu}u^{\nu} +\bigl(P + \frac{b^{2}}{2}\bigr)g^{\mu\nu} - b^{\mu}b^{\nu}\label{eq:s-e} \\
h & = 1 + \epsilon + \frac{P}{\rho_0}\, \\
b^{\mu} & = \frac{1}{\sqrt{4\pi}}B^{\mu}_{(u)}\, \\
B^{0}_{(u)} & = \frac{1}{\alpha}u_i B^{i}\, \\
B^{i}_{(u)} & = \frac{1}{u^{0}}(\frac{B^{i}}{\alpha} + B^{0}_{(u)} u^i)\,.
\end{eqnarray}
Here, $P$ is the fluid pressure, $\rho_0$ is density, $B^i$ is magnetic field, $u^\mu$ is four-velocity, $h$ is the enthalpy, $\epsilon$ is specific internal energy, and $b^2$ is the magnitude of the magnetic vector field squared.  The addition of viscosity modifies the MHD stress-energy tensor by  incorporating the viscous stress tensor
\f
T^{\mu\nu} = (\rho_0 h + b^{2} + Q)u^{\mu}u^{\nu} +\bigl(P + \frac{b^{2}}{2}\bigr)g^{\mu\nu} - b^{\mu}b^{\nu} + \Sigma^{\mu\nu}.\label{eq:s-e-vis}
\ff
Here Q is artificial bulk viscosity and $\Sigma^{\mu\nu}$ is the viscous stress tensor for artificial shear viscosity.  Artificial viscosity is used here as a way of handling shocks although we are working on integrating more advanced HRSC techniques.  Our viscosity terms~\cite{anninos} are described defined below and are only meant to be used when the divergence of the fluid flow is negative.
\begin{eqnarray}
Q = I_{n} ~\Delta l ~\partial_{k} V^{k} ~(k_{q} ~\Delta l ~\partial_{k} V^{k}  - k_{l} ~C_{s}) \\
 \Sigma^{i}_{j} = I_{n}~ \Delta l (k_{q} ~\Delta l ~\partial_{k} V^{k}  - k_{l} ~C_{s}) Sym (\delta_{j} V^{i}  - \frac{\partial_{k} V^{k}}{3}~ \delta^{i}_{j}) \\
 I_{n} = (\rho + \rho \epsilon + (P + Q + b^{2} ) ) ~N \,.
\end{eqnarray}
In these equations $V^{i}$ is the fluid velocity, $C_{s}$ is the local speed of sound, $\Delta l$ is the minimum covariant zone length and~$k_{q}$ and $k_{l}$ are constants multiplying the quadratic and linear contributions, respectively.  The Sym(...) function in the shear viscosity equation is a symmetry operation. 

Dark matter can be added to the system using a two-fluid approach where the stress energy tensor for dark 
matter is  added directly to the stress energy tensor for the magnetofluid, therefore, completing the right-hand side of Einstein$\textquoteright$s equation.  
 
\subsubsection{Initial conditions: Plasma Field} 

In order to get a more accurate picture of the primordial gravitational wave spectrum, we directly simulate the turbulent primordial universe with the most realistic initial conditions possible.  These should include not only a plasma field with a realistic EOS, they should also include elements such as magnetic fields, dark matter and possibly even gravitational waves produced by other sources. 
 
The study will begin at t $>$ $10^{-6}$ seconds after the Big Bang near the beginning of the Hadron epoch, when the primordial plasma field began to look like a relativistic plasma and the strong force could be safely ignored.  At this point the Debye length is about $10^{-16}$ m so even a small computational domain should demonstrate the dynamics of the plasma.  The plasma at this time was composed mainly of electrons, positrons, neutrinos and photons.  Many of the initial conditions at this epoch are well known or can be fairly easily calculated using available literature \cite{islam}.  In addition, given that most cosmological models agree that over 80$\%$ of the matter in the universe is composed of dark matter, our cosmological simulations can also include some non-magnetized ``dark'' fluid.  We can take this into account by adding a second pressureless non-magnetized fluid to our initial plasma field.

The initial matter field will be taken to be homogenous and turbulent.  We introduce turbulence into the system by randomly varying the initial velocities of the fluid elements up to the magnitude of $v_0$, equation (4).  In addition, we will be working to better establish the initial conditions resulting from the first order EWPT.  Based on arguments in previous work~\cite{battaner,Dolgov:2001,grasso,Jedamizik,maroto},  the initial magnetic field during this epoch should be less than or equal to $10^{17}$ G.

\subsubsection{Initial conditions: Spacetime} 

For this computational study the initial spacetime is constructed in such a way as to mimic the conditions present during the Hadron epoch (t  $>$ $10^{-6}$ seconds).  I choose to begin the simulation during the Hadron epoch because at that time the primordial plasma field appeared to look like a relativistic plasma field that could be modeled using a GRMHD code as opposed to a quark gluon plasma field.  At this time the strong force could be safely ignored as electromagnetic effects would dominate the plasma's dynamic motions.  Also, at this time any EWPT would be complete.

Although initial calculations will use a fixed spacetime background, where the background metric is not evolved, we may find it necessary to dynamically model the turbulent/spacetime interactions in order to fully understand the physics of the early universe.  To do this, we need good initial conditions for the curvature of space during this epoch.

The Robertson-Walker (R-W) spacetime metric for a flat universe can be written as 
\f
ds^{2} = -dt^{2} + a^{2}(t)~\bar{g}_{ij} dx^{i} dx^{j},
\ff
where {\it t} is the timelike coordinate, $\bar{g}_{ij}$ is the maximally symmetric three-dimensional space metric, and a({\it t}) the scale factor. For calculating the initial spacetime, conformal time, $\tau = -{1 \over a(t) H(t)}$, is often used instead of cosmic time, $t = t_0 a^2$, to simplify the equations.   Therefore,
\f
ds^{2} = a^{2}(\tau)~g_{\mu\nu} dx^{\mu} dx^{\nu}
\ff
Here, the scale factor and Hubble parameter can be calculated based on the temperature and mass-energy density of the early universe relative to today.  In order to add gravitational waves, the metric $g_{\mu\nu}$ is broken into two components:  $\tilde{g}_{\mu\nu}$, the background metric plus a perturbation, $h_{\mu\nu}$ . The metric can be written as: 
\f
g_{\mu\nu} = \tilde{g}_{\mu\nu} + h_{\mu\nu}.
\ff
The background metric may take any form. It is also assumed that any perturbations are linear. For this study, a R-W metric is presumed. Initial perturbations may or may not be used for numerical experiments within this study.  This metric may involve scalar, ($\phi, \psi$) vector ($\omega_{i}$) and tensor ($h_{ij}$) perturbations, equation (1).  The tensor perturbations are symmetric and transverse-traceless so $\partial_{i} h^{ij} = 0$ and $\delta^{ij} h_{ij} = 0$.  The initial amplitude and spectrum of any of these fluctuations depends on the theory used to explain the generation of perturbations from inflation.  Note that by comparing equations (\ref{eq:adm}) and (\ref{eq:pert}), it can be shown that scalar ($\phi$) and vector ($\omega_{i}$) perturbations can be related to the chosen lapse and shift in the same way that scalar ($\psi$) and tensor ($h_{ij}$) perturbations can relate to the three-metric and extrinsic curvature.   Because the focus here is on tensor perturbations, a geodesic slicing is used so the lapse, $N$, is set to unity and the shift vector,  $N^{i}$, is set to zero.  Since there are no singularities in either the proposed study or the tests described in section 3, geodesic slicing should be sufficient for this work.  Later, if scalar perturbations are included, they can be added by modifying the lapse, as well as the three-metric and extrinsic curvature.  
	
The process of generating tensor perturbations or gravitational waves from quantum fluctuations during inflation is similar to the process of generating scalar or vector perturbations. For example, work by Grishchuk~\cite{Grishchuk:1998qz, Grishchuk:1999et, Grishchuk:2000gh} gives a basis for calculating the spectrum and amplitude of these waves for different slow-roll parameters.  

The possibility of all polarizations are included using $h^{+}_{ij}, h^{\times}_{ij}, h^{L}_{ij}$ and $h^{R}_{ij}$.   Here, $h^{+}_{ij} $ and $h^{\times}_{ij}$ are the plus and cross polarizations respectively, and $h^{L}_{ij}$ and $h^{R}_{ij}$ are the left and right rotating polarizations defined by:
\f
h^{L}_{ij}= {1 \over \sqrt{2}} (h^{+}_{ij} - i h^{\times}_{ij}),   \,  \, \, \, h^{R}_{ij} = {1 \over \sqrt{2}} (h^{+}_{ij}+i h^{\times}_{ij}).
\ff
According to work by Alexander~\cite{alexander}, rotating polarizations should dominate in the early universe and satisfy the equations: 
\f
\Box h^{L}_{ij} = -2 i \theta \dot {h}^{\prime L}_{ij}, \,  \, \Box h^{R}_{ij} = +2 i \theta \dot {h}^{\prime R}_{ij}.
\ff
Here, the dot denotes a time-like derivative with respect to $\tau$ and the prime denotes a spatial derivative along the gravitational wave$\textquoteright$s direction of propagation. If Alexander$\textquoteright$s $\theta$ value is set to zero, the unpolarized gravitational wave signature is recovered~\cite{garrison}.  For this work, rotational polarizations may have the added benefit of introducing extra vorticity into the homogeneous plasma. This may result in an increased magnetic field due to the dynamo effect. During inflation, a non-zero $\theta$ term results in a decrease in the amplitude of $h^L$ and an amplification of $h^R$ and, therefore, cosmological birefringence.  

Initial gravitational wave spectra from sources such as bubble collisions or phase transitions can be added linearly.  Therefore, the initial gravitational wave spectrum can correspond to those predicted by Supersymmetry, Loop Quantum Gravity, String Theory, Deformed Special Relativity, Variable Speed of Light theories, or many other theoretical models in order to provide potential tests of theoretical physics once the system is evolved with the turbulent matter field.

\subsubsection{Numerical Evolutions}
 
The development of a stable and accurate GRMHD code has been a work in progress for the past several years.  We now have a new GRMHD code in the testing and improvement stage.  This code is designed to perform space-time evolutions using either 2nd order finite, 4th order finite or Fourier pseudo-spectral differencing methods with Magnetohydrodynamic and pressure-less matter fields as well as various boundary and gauge conditions.  The code can evolve the matter field independently of the spacetime in cases where a fully dynamic spacetime is not needed so the spacetime metric does not have to be evolved.  

We utilize the Cactus framework to develop this code. We developed an arrangement for Cactus that contains the GRMHD initial data, analysis and evolution thorns.  This code handles the physics while Cactus does the IO and parallelization.  The code is structured so that all the differencing is done outside of the main loops.  This allows us to choose between several differencing techniques such as finite differencing or Fourier spectral differencing at run-time.  We also used the Cactus method of lines routines to supply the time integrators.  The spacetime (LHS of Einstein$\textquoteright$s Equations) can be evolved using a strongly hyperbolic form of the BSSN equations as defined by Brown et al \cite{sarbach}.  The matter field (RHS of EinsteinÕs Equations) is evolved by the form of the GRMHD equations as defined by Duez et al \cite{duez1} with divergence cleaning and artificial viscosity.  Periodic boundary conditions are also used so that the simulation domain can accurately represent a homogenous slice of a much larger universe.

In addition to developing an accurate GRMHD evolution code, it is important to extrapolate the data so that it can be compared to cosmological observations. Gravitational Waves can be calculated directly from the stress-energy tensor using it's quadrupole moments.  By doing this, the spectrum and relative amplitude of primordial gravitational waves created as a result of the turbulence in the matter field can be determined.  Eventually, these results can be compared to stochastic gravitational wave data from GW observatories and observations of the cosmic microwave background.
 
There are many challenges to evolving the numerical code.  First, there are the standard difficulties of dealing with a nonlinear code.  Speed and accuracy are the most important issues.  Also, there are additional challenges because the GRMHD code utilizes a nonlinear Òprimitive variableÓ solver to recover elements of the stress-energy tensor from the MHD evolution variables, shock capturing techniques and a technique called Òdivergence cleaningÓ to maintain physical values for the B-field.  Optimizing and improving these solvers are essential to developing a fast, accurate and stable code.  

Before running the experiments we thoroughly tested the code. The Duez paper~\cite{duez1} suggested four tests of a GRMHD code, however, because of the limited scope of this study I felt that only the following tests are necessary: Gravitational wave-induced MHD waves, Mikowski spacetime MHD tests, such as shock tests and consistency with the standard model of cosmology. I will not include tests of unmagnetized relativistic stars or relativistic Bondi flow in this paper because the spacetime that they are simulating lacks stars and black holes.

This code is being developed and run on a variety of computing resources including: UHCL$\textquoteright$s  Athena cluster, University of Houston$\textquoteright$s  Maxwell cluster and University of Texas$\textquoteright$  Ranger cluster via the XSEDE network.   

\subsubsection{Data Analysis}

The data analysis part of this project will focus around determining the gravitational wave spectrum from the simulation.   A Fourier analysis of the quadrupole moments of the stress-energy tensor should yield the spectrum of gravitational waves produced by the turbulent matter field.  Much of the data analysis work consists of the addition and fine tuning of new analysis routines in the code.

Visualization of Cactus-generated data is done using a variety of open-source software such as VisIt, xgraph, ygraph and gnuplot.  Each requires implementation scripts to be written.  These scripts will tell the visualization program how to read the Cactus-generated data files. Modifications of the data include Fast Fourier Transforms (FFT) for spectral analysis.
 
As these numerical experiments are being performed, the output is analyzed.  The effects of variations in the density, temperature, magnetic field and initial turbulence will be studied in the output data.  A Fourier analysis of the perturbed quadrupole moments will be performed in order to extract the spectrum of gravitational waves.  This spectrum will then be extrapolated to give the current observable values.  The result will be several templates of GW spectra resulting from different initial conditions.  

\section{Testing the Code }
\label{}

The first test that we performed involved generating Alfv\'{e}n and magnetosonic modes by gravitational waves and comparing the results against the semi-analytic solutions from the Duez paper \cite{duez2}.  This semi-analytic solution is only valid for a time much less than the dynamical collapse time of the unperturbed fluid. We began by using the same initial conditions as defined by Duez's general example \cite{duez1}:  

\begin{eqnarray}
h_{+}(t,z) = h_{+ 0}~ sin (kz)~ cos (kt) , ~~ h_{\times }(t,z) = h_{\times  0}~ sin (kz)~ cos (kt) , \\
P(0,z) = 1.29 \times  10^{-9} ,~~          \rho_0(0,z) = 2.78 \times 10^{-9}, \\
v^{i}(0,z) = 0 ,~~           B^{i}(0,z) =(1.09, 8.26, 14.4) \times  10^{-5} , \\
h_{+ 0} = h_{\times  0} = 1.18 \times  10^{-4} ,
\end{eqnarray}

where k is the wave number and we assume the fluid is unperturbed at t = 0.  Results are shown in Figure 1.  The test was performed using second order, fourth order and Fourier spectral differencing as a one dimensional problem.  The results shown used 200 grid points for the second order differencing, 50 grid points for the fourth order differencing and 32 grid points for the spectral differencing.  Additional tests were performed where the initial conditions were varied.  For every variation the test proved successful, the analytic and numerical results proved almost identical to within a few percent.  Runs were also conducted with 100, 25 and 16 grid points for the second order, fourth order and spectral differencing respectively, in order to test for convergence.  As shown in Figure 1, the main source of errors in this test were phase errors between the analytic and numerical solutions.  This made calculating convergence difficult.  We were able to calculate convergence of the results at each time by dividing the L2 norm of the errors of the low resolution runs by that of the high resolution runs.  The result was an oscillating pattern with a mean, after 5 crossing times, of around 2.4 for the 2nd order finite differencing, 4.67 for the 4th order finite differencing and 2.7 for the Fourier Spectral differencing.  A major factor effecting the convergence rate may be the larger time steps taken in the low resolution runs.  Although we cannot show that the overall convergence rate matches that of the differencing method, we can show that the code does converge for each differencing method.  

\begin{figure}  
\begin{center}
\includegraphics[height=5.0in,width=6.5in,angle=0]{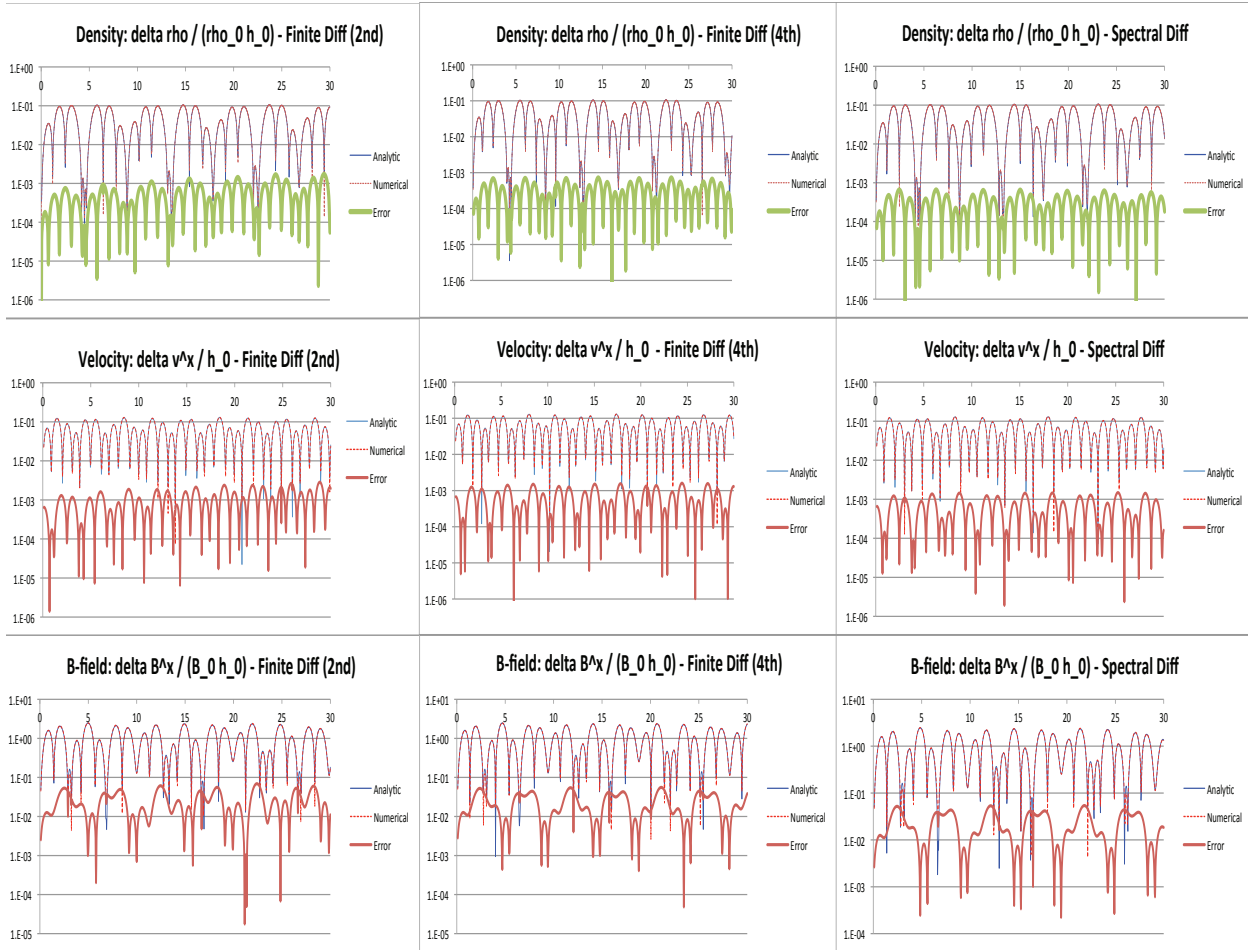}
\caption{\small \sl Gravitational Wave-Induced MHD waves test results.  The left column used 2nd order finite differencing, the middle column used 4th order finite differencing and the right column used Fourier spectral differencing.  Density is displayed here by calculating $\delta \rho / ( \rho h_{0} )$, Velocity is displayed here by calculating $\delta v^x / h_{0}$ and B-field is displayed here by calculating $\delta B^{x} / ( B_{0} h_{0} )$ as in the Duez paper \cite{duez1}. Time is shown on the x axis.  Because the physical size of the grid is 2 units, a time of 30 units represents 15 crossing times.  \label{fig:waves1}}  
\end{center}
\end{figure} 

For our second test we utilized a FRW space-time that contained parameters (temperature, energy density, Hubble parameter and scale factor) based with those accepted for the universe at t = $10^{-6}$ s. The goal of the test was to determine if it evolved consistently with the Friedmann equations.  We set the initial scale factor a = 6.6 $\times$ $10^{-14}$, the initial Hubble parameter H = 1.4 $\times$ $10^{26}$ km/s/Mpc, the initial temperature T = 4.1 $\times$ $10^{13}$ K and the initial energy density $\rho_{0} \epsilon$ = 1.1 $\times$ $10^{23}$ J/$m^3$ .  The system evolved until around t = 86 s and the final value of these parameters were found at the end of the simulation.  The code's calculated change in each parameter was then compared to the values predicted from the Friedmann equations in order to determine how well the code's results matched the analytic solution.  The final time, 86s, corresponded to physical age of the simulated universe after seven days of computing time, and was not chosen to have any particular relationship to the physical size of the domain.  There were no structures or inhomogeneities in the system so spacial resolution was not important in this test.  This test proved successful with the errors of less than one percent as shown in Table 1.

\begin{table*}
\caption{Consistency of Standard Model of Cosmology parameters test results.\label{table:cosmo}}
\begin{tabular}{c  c  c  c c c}
\hline \hline
 & t & T & $\rho \epsilon$ & H & a \\ 
\hline \hline
Initial & $10^{-6}$ & $4.120 \times 10^{13}$ & $1.091 \times 10^{23}$ & $2.401 \times 10^{26}$ & 1.0 \\
Final & $8.629 \times 10^{1}$ & $4.446 \times 10^{9}$ & $1.480 \times 10^{7}$ & $2.792 \times 10^{18}$ & $9.289 \times 10^{3}$  \\
Final/Initial & $8.629 \times 10^{7}$ & $1.079 \times 10^{-4}$ & $ 1.357 \times 10^{-16}$ & $1.163 \times 10^{-8}$ & $9.289 \times 10^{3}$  \\
\hline
t${}^{\rm a}$ & $8.629 \times 10^{7}$ & $8.586 \times 10^{7}$ & $8.586 \times 10^{7}$ & $8.601 \times 10^{7}$ & $8.629 \times 10^{7}$ \\
T${}^{\rm a}$ & $1.077 \times 10^{-4}$ & $1.079 \times 10^{-4}$ & $1.079 \times 10^{-4}$ & $1.078 \times 10^{-4}$ & $1.077 \times 10^{-4}$ \\
$\rho \epsilon$${}^{\rm a}$ & $1.343 \times 10^{-16}$ & $1.357 \times 10^{-16}$ & $1.357 \times 10^{-16}$ & $1.352 \times 10^{-16}$ & $1.343 \times 10^{-16}$ \\
H${}^{\rm a}$ & $1.159 \times 10^{-8}$ & $1.165 \times 10^{-8}$ & $1.165 \times 10^{-8}$ & $1.163 \times 10^{-8}$ & $1.159 \times 10^{-8}$ \\
a${}^{\rm a}$ & $9.289 \times 10^{3}$ & $9.266 \times 10^{3}$ & $9.266 \times 10^{3}$ & $9.274 \times 10^{3}$ & $9.289 \times 10^{3}$ \\
\hline
t${}^{\rm b}$ & 0.0 & $4.338 \times 10^{5}$ & $4.338 \times 10^{7}$ & $2.824 \times 10^{5}$ & -$2.980 \times 10^{-7}$ \\
T${}^{\rm b}$ & $2.716 \times 10^{-7}$ & 0.0 & -$5.833 \times 10^{-16}$ & $9.506 \times 10^{-8}$ & $2.716 \times 10^{-7}$ \\
$\rho \epsilon$${}^{\rm b}$ & $1.361 \times 10^{-18}$ & $2.933 \times 10^{-27}$ & 0.0 & $4.773 \times 10^{-19}$ & $1.361 \times 10^{-18}$ \\
H${}^{\rm b}$ & $3.804 \times 10^{-11}$ & -$2.051 \times 10^{-11}$ & -$2.051 \times 10^{-11}$ & 0.0 & $3.804 \times 10^{-11}$ \\
a${}^{\rm b}$ & $1.637 \times 10^{-11}$ & $2.338 \times 10^{1}$ & $2.338 \times 10^{1}$ & $1.521 \times 10^{1}$ & 0.0 \\
\hline
t${}^{\rm c}$ & $3.351 \times 10^{-3}$  \\
T${}^{\rm c}$ & & $1.678 \times 10^{-3}$  \\
$\rho \epsilon$${}^{\rm c}$ & & & $6.686 \times 10^{-3}$  \\
H${}^{\rm c}$ & & & & $3.357 \times 10^{-3}$  \\
a${}^{\rm c}$ & & & & & $1.678 \times 10^{-3}$  \\
\hline \hline
\end{tabular}
\vskip 12pt
\begin{minipage}{12cm}
\raggedright
${}^{\rm a}$ Final/Initial for each column parameter calculated based on the Friedman equations\\
${}^{\rm b}$ ratio predicted by Friedman equations (analytic) - ratio calculated by the GRMHD code (numerical) \\
${}^{\rm c}$ Average Error:  (numerical - analytic) / numerical averaged among all nonzero columns \\
\end{minipage}
\label{tab:1Ctests}
\end{table*}

At this point we are prepared to add standing gravitational waves with a spectrum consistent to Grishchuk$'$s predictions ~\cite{Grishchuk:1998qz, Grishchuk:1999et, Grishchuk:2000gh}.   The gravitational waves were given random phases in order to avoid large nodes and anti-nodes.  We then ran simulations with space-time perturbations and large ($10^{17}$ Gauss) magnetic fields.  These simulations showed that space-time perturbations and magnetic fields had no significant impact on the expansion rate of the space-time and therefore adding them should still result in a simulation consistent with cosmological theory to within the same error as found in Table 1.  

Finally, we conducted shock tests using similar initial conditions to Duez \cite{duez1} and Komissarov  \cite{k97} as shown in Table 2.  The results shown are based on runs utilizing the second order finite differencing method.  The fourth order finite differencing method gives similar results, as expected, although the fourth order tests were conducted using a lower resolution grid.  These tests could not be completed with our Fourier spectral differencing method because the suggested shock tests are not periodic in nature.  Future work may involve testing all three differencing methods using a periodic shock test but the current results suggest that this may not be a worthwhile effort until HRSC techniques can be incorporated into the code.  

\begin{table*}
\caption{Initial states for one-dimensional MHD tests.${}^{\rm a}$}
\begin{tabular}{c  c  c  c c}
\hline \hline
 Test & Left state & Right State & Grid & $t_{\rm final}$ \\ 
\hline \hline
  Fast Shock & $u^i=(25.0,0.0,0.0)$ & $u^i=(1.091,0.3923,0.00)$ & n = 40 & 2.5 \\
 ($\mu=0.2{}^{\rm b}$) & $B^i/\sqrt{4\pi}=(20.0,25.02,0.0)$ & 
 $B^i/\sqrt{4\pi}=(20.0,49.0,0.0)$ &  \\
  & $P=1.0$, $\rho_0=1.0$ & $P=367.5$, $\rho_0=25.48$ & \\
\hline
  Slow Shock & $u^i=(1.53,0.0,0.0)$ & $u^i=(0.9571,-0.6822,0.00)$ & n = 200 & 2.0 \\
 ($\mu=0.5{}^{\rm b}$) & $B^i/\sqrt{4\pi}=(10.0,18.28,0.0)$ & 
 $B^i/\sqrt{4\pi}=(10.0,14.49,0.0)$ & \\
  & $P=10.0$, $\rho_0=1.0$ & $P=55.36$, $\rho_0=3.323$ & \\
\hline
  Switch-off Fast & $u^i=(-2.0,0.0,0.0)$ & $u^i=(-0.212,-0.590,0.0)$ & n = 150 & 1.0 \\
  Rarefaction & $B^i/\sqrt{4\pi}=(2.0,0.0,0.0)$ & 
  $B^i/\sqrt{4\pi}=(2.0,4.71,0.0)$ & \\
  & $P=1.0$, $\rho_0=0.1$ & $P=10.0$, $\rho_0=0.562$ & \\
\hline
  Switch-on Slow & $u^i=(-0.765,-1.386,0.0)$ & $u^i=(0.0,0.0,0.0)$ & n = 150 & 2.0 \\
  Rarefaction & $B^i/\sqrt{4\pi}=(1.0,1.022,0.0)$ & 
 $B^i/\sqrt{4\pi}=(1.0,0.0,0.0)$ & \\
 & $P=0.1$, $\rho_0=1.78\times 10^{-3}$ & $P=1.0$, $\rho_0=0.01$ & \\
 \hline
Alfv\'en Wave${}^{\rm c}$ & $u^i=(0.0,0.0,0.0)$  & 
$u^i=(3.70,5.76,0.00)$ & n = 200 & 2.0 \\
($\mu=0.626{}^{\rm b}$) & $B^i/\sqrt{4\pi}=(3.0,3.0,0.0)$ & 
$B^i/\sqrt{4\pi}=(3.0,-6.857,0.0)$ & \\
 & $P=1.0$, $\rho_0=1.0$ & $P=1.0$, $\rho_0=1.0$ & \\
\hline
  Shock Tube 1 & $u^i=(0.0,0.0,0.0)$ & $u^i=(0.0,0.0,0.0)$ & n = 200 & 1.0 \\
  & $B^i/\sqrt{4\pi}=(1.0,0.0,0.0)$ & $B^i/\sqrt{4\pi}=(1.0,0.0,0.0)$ & \\
  & $P=1000.0$, $\rho_0=1.0$ & $P=1.0$, $\rho_0=0.1$ & \\
\hline
  Shock Tube 2 & $u^i=(0.0,0.0,0.0)$ & $u^i=(0.0,0.0,0.0)$ & n = 200 & 1.0 \\
 & $B^i/\sqrt{4\pi}=(0.0,20.0,0.0)$ & $B^i/\sqrt{4\pi}=(0.0,0.0,0.0)$ & \\
 & $P=30.0$, $\rho_0=1.0$ & $P=1.0$, $\rho_0=0.1$ & \\
\hline
  Collision & $u^i=(5.0,0.0,0.0)$ & $u^i=(-5.0,0.0,0.0)$ & n = 200 & 1.22 \\
  & $B^i/\sqrt{4\pi}=(10.0,10.0,0.0)$ & $B^i/\sqrt{4\pi}=(10.0,-10.0,0.0)$ & \\
  & $P=1.0$, $\rho_0=1.0$ & $P=1.0$, $\rho_0=1.0$ & \\
\hline \hline
\end{tabular}
\vskip 12pt
\begin{minipage}{12cm}
\raggedright
${}^{\rm a}$ {In all cases, the gas satisfies the $\Gamma$-law EOS with 
$\Gamma = 4/3$. For the first 7 tests, the left state refers to $x<0$ 
and the right state, $x>0$.} \\
${}^{\rm b}$ {$\mu$ is the speed at which the wave travels} \\
${}^{\rm c}$ {For the nonlinear Alfv\'en wave, the left and right 
states are joined by a continuous function separated by 0.5 units.} \\
\end{minipage}
\label{tab:1Dtests}
\end{table*}

\begin{figure}  
\begin{center}
\centering
 \includegraphics[height=5.0in,width=6.5in,angle=0]{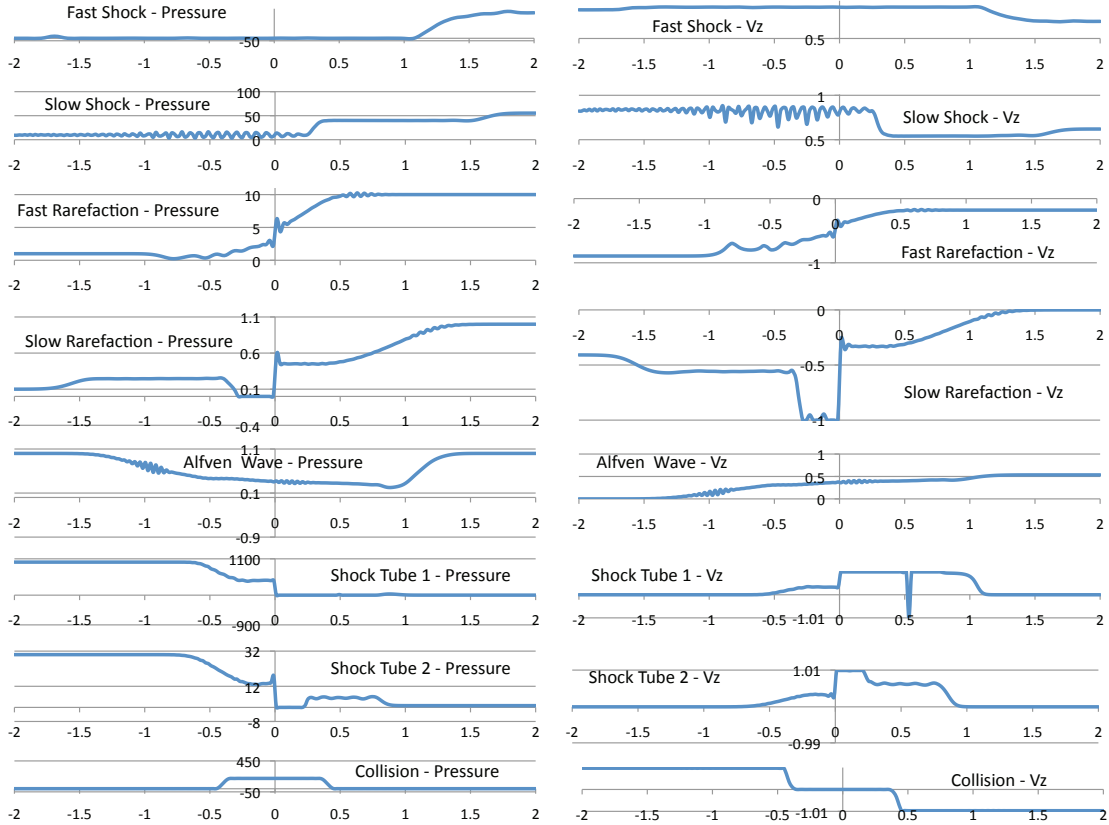}
\caption{\small \sl Shock Test Results shown using second order finite differencing and artificial viscosity at $t_{final}$ time. The pressure profiles are shown on the left and the z component of velocity profiles are shown on the right.\label{fig:cosmo}} 
\end{center}
\end{figure} 

Overall, the artificial viscosity methods used in this code seemed to produce less accurate results than the High Resolution Shock Capturing methods used by Duez \cite{duez1} and Komissarov \cite{k97}.   {\it Fast and Slow Shocks}: The shock fronts for both of these cases was more distorted than the results of Duez \cite{duez1}and Komissarov \cite{k97}.  Also, the Fast Shock appeared to move slower than the Slow Shock which doesn't seem to agree with the standard results.  {\it Switch-on/off Rarefaction}: While the Switch-off (Fast Rarefaction) seemed to agree with the published results the Switch-on (Slow Rarefaction) appeared distorted at the shock front.  {\it Alfv\'en Wave}: Our Alfv\'en Wave results seemed to agree with the published results except for the extra dip for $z < 0$.  {\it Shock Tubes 1 and 2}: Shock Tube tests also produced distorted shock fronts, particularly for $z > 0$. {\it Collision}: The Collision test produced the best results when compared to the established results. 

\section{Preliminary Results}
\label{}

In order to present a relevant proof of concept on the use of GRMHD in cosmology, we evolve a turbulent plasma field, with conditions similar to the universe when it was $10^{-6}$ s old.  We use similar conditions as outlined in section 2.2.3 and the Consistency of Standard Model of Cosmology parameters test with an initial uniform magnetic field of $10^{15}$ G.  No initial gravitational waves or dark matter was included in the system.  We also introduced turbulence with a random initial velocity of 0.65.   

\begin{figure}  
\begin{center}
\centering
 \includegraphics[height=5.0in,width=6.5in,angle=0]{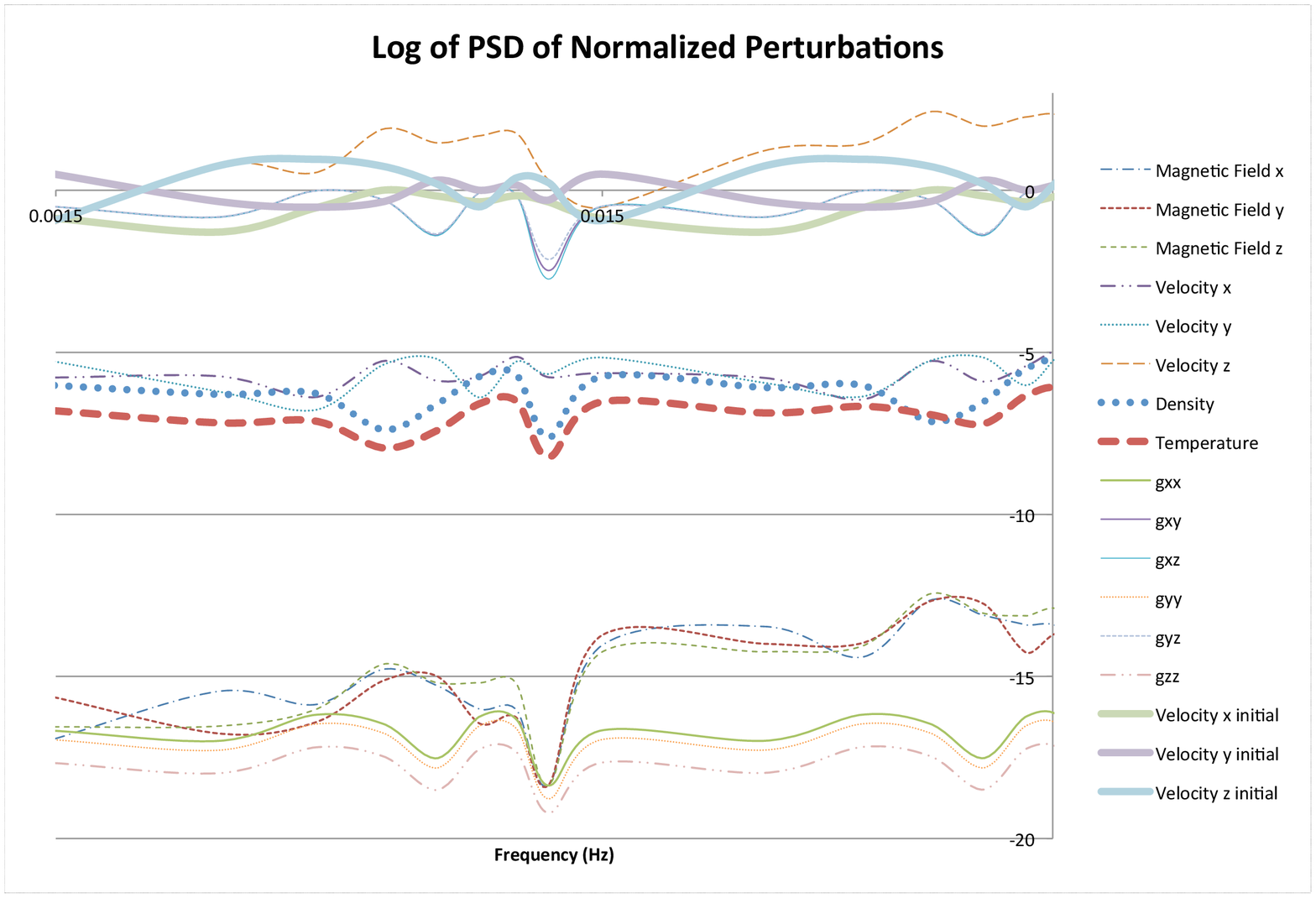}
\caption{\small \sl Logrithms of the Power Spectral Densities of several quantities after a turbulent relativistic plasma was evolved using the GRMHD code from $10^{-6}$ s to around 3.5 $\times 10^{-5}$ s.  The perturbations were normalized using the mean amplitude of each quantity before the PSD was calculated. \label{fig:PSD}} 
\end{center}
\end{figure} 

The data presented in Figure 3 correspond to evolving the initial conditions stated above to around 3.5 $\times 10^{-5}$ s.  Before calculating the PSDs of each of the quantities, they where normalized by dividing the perturbation amplitude by the mean value of the quantity.  The results where plotted Logrithmically for a frequency range of 0.0015 Hz to 0.12 Hz.  Although the simulation was run to 3.5 $\times 10^{-5}$ s, the normalized PSDs didn't seem to vary much after about a hundred iterations.  The normalized PSDs of the relevant parameters were plotted on a logarithmic scale for a frequency range relevant to a potential space based gravitational wave interferometer such as eLISA.  The results show that for strong uniform initial magnetic fields, noticeable perturbations are generated in the space-time metric, density, temperature and magnetic field terms which are different from the perturbations in the velocity field.  Perturbations in the metric are the most interesting of these results because they correspond to gravitational waves.  These where vanishingly small for magnetic fields less than or equal to $10^{12}$ G.  Much more work is needed to more fully understand the dynamics of the interaction between GRMHD turbulence and gravitational waves but the results so far clearly show that gravitational wave generation from primordial turbulence is possible.

\section{Discussion}
\label{}

We now have several parameters to work with in developing numerical experiments.  Assuming the expansion properties of the background spacetime and composition of the matter field are fixed, we can alter the metric perturbations (scalar and tensor), initial magnetic field strength and the turbulence of the initial matter field.  By altering these properties, we can perform a variety of numerical experiments to determine the effects of scalar perturbations and gravitational waves on structure formation, limits on primordial magnetic fields, properties of gravitational waves formed by a turbulent plasma, the dynamics of a turbulent plasma in an expanding universe and other interesting scientific properties.

One of the most interesting of these experiments involves the interaction between gravitational waves and the primordial plasma.  Work by Duez \cite{duez2}, showed that gravitational waves can induce oscillatory modes in a plasma.  According to Shebalin \cite{shebalin}, large-scale coherent structures grow naturally out of MHD turbulence.  Here, structure is defined as strengthening magnetic fields, permanent density and temperature variations and secondary relic gravitational waves.  One can assume that space-time perturbations in the early universe (sometime after t = $10^{-6}$ seconds) interacted with the primordial plasma and resulted in Alfv\'{e}n and magnetosonic modes \cite{duez2}.  These modes then interacted dynamically, possibly resulting in turbulence and structure formation \cite{shebalin}.  Using the techniques of numerical relativity, we can test this assumption.

\ack
The author would like to acknowledge support from the Institute for Space Systems Operations and the University of Houston Clear LakeÕs Faculty Research and Support Funds.  The author would also like to acknowledge the masterÕs degree students who worked on several aspects of this project over the past several years: Cindi Ballard, David Chow, John Hamilton, Tom Smith, Kevin Depaula, Rafael de la Torre, Marlo Graves, Chris Greenfield and Paul Smith.  In addition, I would like to thank Drs. John Shebalin,  Leonard Kisslinger, Tina Kahniashvili, Bernard Kelly and Samina Masood for many useful conversations and helpful suggestions.


\end{document}